\documentstyle[12pt]{article}
\begin{document}
\title{The Nature of Space Time}
\author{B.G. Sidharth\\
Centre for Applicable Mathematics \& Computer Sciences\\
B.M. Birla Science Centre, Adarsh Nagar, Hyderabad - 500 063 (India)}
\date{}
\footnotetext{Email:birlasc@hd1.vsnl.net.in}
\maketitle
\begin{abstract}
We first examine the approximation involved in the conventional differentiable
spacetime manifold. We then analyse how, going beyond this approximation, we
reach the non commutative spacetime of recent approaches. It is shown that
this provides the rationale for El Naschie's transfinite Cantorian spacetime.
The nature and form of the consequent Generalized Uncertainity Principle is also
briefly investigated.
\end{abstract}
\section{Introduction}
All of Classical Physics and Quantum Theory, is based on the Minkowski
spacetime, as for example in the case of Quantum Field Theory, or Reimannian
spacetime as in the case of General Relativity. In the non relativistic
theories, Newtonian spacetime, is used, which is a special case of Minkowskian
spacetime. But in all these cases the common denominator is that we are
dealing with a differentiable manifold.\\
This breaks down however in Quantum Gravity, String Theory and more recent
approaches, be it at the Planck scale, or at the Compton scale\cite{r1,r2,r3,r4}.
The underlying reason for this breakdown of a differentiable spacetime manifold
is the Uncertainty Principle-- as we go down to arbitrarily small spacetime
intervals, we encounter arbitrarily large energy momenta. As Wheeler put
it\cite{r5}, "no prediction of spacetime, therefore no meaning for spacetime
is the verdict of the Quantum Principle. That object which is central to all
of Classical General Relativity, the four dimensional spacetime geometry,
simply does not exist, except in a classical approximation." Before proceeding
to analyse the nature of spacetime beyond the classical approximation, let
us first analyse briefly the nature of classical spacetime itself.
\section{The "Classical" Approximation}
We can get an insight into the nature of the usual spacetime by considering
the well known formulation of Quantum Theory in terms of stochastic
processes \cite{r6,r7,r8,r9}. This will also facilitate subsequent considerations.\\
In the stochastic or Nelsonian theory, we deal with a double Weiner process
which leads to a complex velocity $V-\imath U$. It is this complex velocity that leads to
Quantum Theory from the usual diffusion theory (Cf.\cite{r7} for details).\\
To see this in a simple way, let us write the usual diffusion equation as
\begin{equation}
\Delta x \cdot \Delta x = \frac{h}{m}\Delta t \equiv \nu \Delta t\label{e1}
\end{equation}
Equation (\ref{e1}) can be rewritten as the usual Quantum Mechanical relation,
\begin{equation}
m\frac{\Delta x}{\Delta t} \cdot \Delta x = h = \Delta p \cdot \Delta x\label{e2}
\end{equation}
We are dealing here, with phenomena within the Compton or de Broglie wavelength.\\
We now treat the diffusion constant $\nu$ to be very small, but
non vanishing. That is, we consider the semi classical case. This is because,
a purely classical description, does not provide any insight.\\
It is well known that in this situation we can use the WKB approximation. In this
case the right hand side of the representation of the Nelsonian wave
function,
$$\psi = \sqrt{\rho} e^{\imath /\hbar S}$$
goes over to, in the one dimensional case, for simplicity,
$$(p_x) ^{-\frac{1}{2}} e^{\frac{1}{h}} \int p(x)dx$$
so that we have, on comparison,
\begin{equation}
\rho = \frac{1}{p_x}\label{e3}
\end{equation}
$\rho$ being the probability density. In this case the condition $U \approx 0$,
that is, the velocity potential becoming real, implies
\begin{equation}
\nu \cdot \nabla ln (\sqrt{\rho}) \approx 0\label{e4}
\end{equation}
This semi classical analysis suggests that $\sqrt{\rho}$ is a slowly varying
function of $x$, infact each of the factors on the left side of (\ref{e4}) would
be $\sim 0(h)$, so that the left side is $\sim 0(h^2)$ (which is being neglected).
Then from (\ref{e3}) we conclude that $p_x$ is independent of $x$, or is a slowly
varying function of $x$. The equation of continuity now gives
$$\frac{\partial \rho}{\partial t} + \vec \nabla (\rho \vec v)  = \frac{\partial \rho}
{\partial t} = 0$$
That is the probability density $\rho$ is independent or nearly so, not only of
$x$, but also of $t$. We are thus in a stationary and homogenous scenario. This
is strictly speaking, possible only in a single particle universe, or for a completely
isolated particle, without any effect of the environment. Under these circumstances
we have the various conservation laws and the time reversible theory, all this
taken over into Quantum Mechanics as well. This is an approximation valid for
small, incremental changes, as indeed is implicit in the concept of a
differentiable space time manifold.\\
Infact the well known displacement operators of Quantum Theory which define
the energy momentum operators are legitimate  and further the energy and momenta
are on the same footing only under this approximation\cite{r10}.\\
We would now like to point out the well known close similarity between the
Nelsonian formulation mentioned above (Cf.(\ref{e1}) and (\ref{e2})
and the hydrodynamical formulation for
Quantum Mechanics, which also leads to identical equations on writing the
wave function as above. These two approaches were reconciled by considering
quantized vortices at the Compton scale (Cf.\cite{r11,r9}). To proceed further,
we start with the Schrodinger equation
\begin{equation}
\imath \hbar \frac{\partial \psi}{\partial t} = - \frac{\hbar^2}{2m} \nabla^2
\psi + V \psi\label{e5}
\end{equation}
Remembering that for momentum eigen states we have, for simplicity, for one
dimension
\begin{equation}
\frac{\hbar}{\imath} \frac{\partial}{\partial x} \psi = p\psi\label{e6}
\end{equation}
where $p$ is the momentum or $p/m$ is the velocity $v$, we take the derivative
with respect to $x$ (or $\vec x$) of both sides of (\ref{e5}) to obtain,
on using (\ref{e6}),
\begin{equation}
\imath \hbar \frac{\partial (v\psi )}{\partial t} = - \frac{\hbar^2}{2m}
\nabla^2 (v \psi) + \frac{\partial V}{\partial x} \psi + Vv\psi\label{e7}
\end{equation}
We would like to compare (\ref{e7}) with the well known equation for the
velocity in hydrodynamics\cite{r12}, following from the Navier-Stokes equation,
\begin{equation}
\rho \frac{\partial v}{\partial t} = -\nabla p - \rho \alpha T g + \mu
\nabla^2 v\label{e8}
\end{equation}
In (\ref{e8}) $v$ is a small perturbational velocity in otherwise stationary
flow between parallel plates separated by a distance $d, p$ is a small pressure,
$\rho$ is the density of the fluid $T$ is the temperature proportional to
$Q(z)v,\mu$ is the Navier-stokes coefficient and $\alpha$ is the coefficient
of volume expansion with temperature. Also required would be
$$\beta \equiv \frac{\Delta T}{d}.$$
$v$ itself is given by
\begin{equation}
v_z = W(z)exp (\sigma t + \imath k_xx + \imath k_y y),\label{e9}
\end{equation}
$z$ being the coordinate perpendicular to the fluid flow.\\
We can now see the parallel between equations (\ref{e7}) and (\ref{e8}). To
verify the identification we would require that the dimensionless Rayleigh number
$$R = \frac{\alpha \beta g d^4}{\kappa \nu}$$
should have an analogue in (\ref{e7}) which is dimensionless, $\kappa , \nu$
being the thermometric conductivity and viscocity.\\
Remembering that
$$\nu \sim \frac{h}{m}$$
and
$$d \sim \lambda$$
where $\lambda$ is the Compton wavelength in the above theory (Cf.\cite{r9} for
details) and further we have
\begin{equation}
\rho \propto f(z)g = V\label{e10}
\end{equation}
for the identification between the hydrostatic energy and the energy $V$ of
Quantum Mechanics, it is easy using (\ref{e10}) and earlier relations to show that the analogue of $R$ is
\begin{equation}
(c^2/\lambda^2) \cdot \lambda^4 \cdot (m/h)^2\label{e11}
\end{equation}
The expression (\ref{e11}) indeed is dimensionless and of order $1$. Thus the mathematical identification
is complete.\\
Before proceeding, let us look at the physical significance of the above
considerations (Cf.\cite{r13} for a graphic description.) Under conditions of stationery flow, when the
diifference in the temperature between the two plates is negligible there is
total translational symmetry, as in the case of the displacement operators of
Quantum Theory. But when there is a small perturbation in the velocity (or
equivalently the temperature difference), then beyond a critical value the
stationarity and homogeneity of the fluid is disrupted, or the symmetry is broken
and we have the phenomena of the formation of Benard cells, which are
convective vortices and can be counted. This infact is the "birth" of space
(Cf.\cite{r13} for a detailed description).\\
In the context of the above identification, the Benard cells would correspond
to the formation of quantized vortices, which latter had been discussed in
detail in the literature (Cf.\cite{r9} and \cite{r14}). This transition would
correspond to the "formation" of spacetime. As discussed in detail in \cite{r9}
these quantized vortices can be identified with elementary particles, in
particular the electrons. Interestingly, Einstein himself considered electrons
as condensates from a background electromagnetic field\cite{r15}.\\
However in order to demonstrate that the above quantized vortex formation is
not a mere mathematical analogy, we have to show that the critical value of
the wave number $k$ in the expression for the velocity in the hydrodynamical
flow (\ref{e9}) is the same as the critical value of the quantized vortex length. In
terms of the dimensionless wave number $k' = k/d$, this critical value is
given by\cite{r12}
$$k'_c \sim 1$$
In the case of the quantized vortices at the Compton scale $l$, remembering
that $d$ is identified with $l$ itself we have,
$$l'_c (\equiv) k'_c \sim 1,$$
exactly as required.\\
In this connection it may be mentioned that due to fluctuations in the
Zero Point Field or the Quantum vaccuum, there would be fluctuations in the
metric given by\cite{r5}
$$\Delta g \sim l_P/l$$
where $l_P$ is the Planck length $\sim 10^{-33}cms$ and $l$ is a small interval
under consideration. At the same time the fluctuation in the curvature of
space would be given by
$$\Delta R \sim l_P/l^3$$
Normally these fluctuations are extremely small but as discussed in detail
elsewhere\cite{r16}, this would imply that at the Compton scale of a typical
elementary particle $l \sim 10^{-11}cms$, the fluctuation in the curvature
would be $\sim 1$. This is symptomatic of the formation of what we have
termed above as quantized vortices.\\
Further if a typical time interval between the formation of such quantized
vortices which are the analogues of the Benard cells is $\tau$, in this
case the Compton time, then as in the theory of the Brownian Random Walk\cite{r17},
the mean time extent would be given by
\begin{equation}
T \sim \sqrt{N}\tau\label{e12}
\end{equation}
where $N$ is the number of such quantized vortices or elementary particles
(Cf.also \cite{r9,r11}). It is quite remarkable that the equation (\ref{e12})
holds good for the universe itself because $T$ the age of the universe $\sim 10^{17}
secs$ and $N$ the number of elementary particles $\sim 10^{80}, \tau$
being the Compton time $\sim 10^{-23} secs$. Interestingly, this nature of
time would automatically make it irreversible, unlike the conventional model in
which time is reversible.\\
It may be mentioned that an equation similar to (\ref{e12}) can be deduced by the same arguments for
space extension also and this time we get the well known Eddington formula
viz.,
\begin{equation}
R \sim \sqrt{N} l\label{e13}
\end{equation}
where $R$ is the radius of the universe and $l$ is the Compton wavelength.
Further starting from (\ref{e12}) one can work out a whole scheme of what
may be called fluctuational cosmology, in which not just the Eddington
formula (\ref{e13}) above, but also all the other supposedly mysterious and
inexplicable large number relations of Dirac and the Weinberg formula relating
the mass of the pion to the Hubble Constant can be deduced theoretically.
Furthermore, this cosmology predicts an ever expanding and accelerating
universe, as is now recognised to be the case (Cf.\cite{r18,r9} for details).\\
Once we recognize the minimum space time extensions, then we immediately
are lead to an underlying non commutative geometry given by\cite{r9},
\begin{equation}
[x,y] = 0(l^2),[x,p_x] = \imath \hbar [1+0(l^2)], [t,E]=\imath \hbar
[1 + 0(\tau^2)\label{e14}
\end{equation}
As was shown a long time ago, relations like (\ref{e14}) are Lorentz invariant.
At this stage we recognise the nature of spacetime as given by (\ref{e14})
in contrast to the stationary and homogeneous spacetime discussed earlier.
Witten\cite{r19} has called this Fermionic spacetime as contrasted to the
usual spacetime, which he terms Bosonic. Indeed one could show the origins
of the Dirac equation of the electron from (\ref{e14}). We could also argue
that (\ref{e14}) provides the long sought after reconciliation between
electromagnetism and gravitation\cite{r20,r21}.\\
The usual differentiable spacetime geometry can be obtained from (\ref{e14})
if $l^2$ is neglected, and this is the approximation that has been implicit.
\section{Cantorian Spacetime and Metric}
Thus spacetime is a collection of such cells or elementary particles very
much in the spirit of El Naschie's Cantorian spacetime\cite{r22,r23,r24}.
As pointed out earlier, this spacetime emerges from a homogeneous stationary
non spacetime when the symmetry is broken, through random processes. The
question that comes up then is, what is the metric which we use? This has been
touched upon earlier, and we will examine it again.\\
We first makes a few preliminary remarks. When we talk of a metric or the
distance between two "points" or "particles", a concept that is implicit is that
of topological "nearness" - we require an underpinning of a suitably large
number of "open" sets\cite{r25}. Let us now abandon the absolute or background
space time and consider, for simplicity, a universe (or set) that consists solely
of two particles. The question of the distance between these particles (quite
apart from the question of the observer) becomes meaningless. Indeed, this is
so for a universe consisting of a finite number of particles. For, we could
isolate any two of them, and the distance between them would have no meaning.
We can intuitiively appreciate that we would infact need distances of intermediate
or more generally, other points.\\
In earlier work\cite{r26}, motivated by physical considerations we had considered
a series of nested sets or neighbourhoods which were countable and also whose
union was a complete Hausdorff space. The Urysohn Theorem was then invoked and
it was shown that the space of the subsets was metrizable.
The argument went something like this.\\
In the light of the above remarks, the concepts of open sets, connectedness and
the like reenter in which case such an isolation of two points would not be possible.\\
More formally let us define a neighbourhood of a particle (or point or element) $A$ of a
set of particles as a subset which contains $A$ and atleast one other distinct
element. Now, given two particles (or points) or sets of points $A$ and $B$, let us consider a
neighbourhood containing both of them, $n(A,B)$ say. We require a non empty
set containing atleast one of $A$ and $B$ and atleast one other particle $C$, such
that $n(A,B) \subset n(A,C)$, and so on. Strictly, this "nested" sequence
should not terminate. For, if it does, then we end up with a set $n(A,P)$ consisting
of two isolated "particles" or points, and the "distance" $d(A,P)$ is meaningless.\\
We now assume the following property\cite{r26}: Given two distinct elements (or
even subsets) $A$ and $B$, there is a neighbourhood $N_{A_1}$ such that $A$ belongs
to $N_{A_1}$, $B$ does not belong to $N_{A_1}$ and also given any $N_{A_1}$, there
exists a neighbourhood $N_{A_\frac{1}{2}}$ such that $A \subset N_{A_\frac{1}{2}}
\subset N_{A_1}$, that is there exists an infinite topological closeness.\\
From here, as in the derivation of Urysohn's lemma\cite{r25}, we could define a
mapping $f$ such that $f(A) = 0$ and $f(B) = 1$ and which takes on all intermediate
values. We could now define a metric, $d(A,B) = |f(A) - f(B)|$. We could
easily verify that this satisfies the properties of a metric.\\
With the same motivation we will now deduce a similar result, but with different
conditions. In the sequel, by a subset we will mean a proper subset, which is also
non null, unless specifically mentioned to be so. We will also consider
Borel sets, that is the set itself (and its subsets) has a countable covering
with subsets. We then follow a pattern similar to that of a Cantor ternary set
\cite{r25,r27}. So starting with the set $N$ we consider a subset $N_1$ which
is one of the members of the covering of $N$ and iterate this process so that
$N_{12}$ denotes a subset belonging to the covering of $N_1$ and so on.\\
We note that each element of $N$ would be contained in one of the series of subsets of a
sub cover. For, if we consider the case where the element $p$ belongs to
some $N_{12\cdots m}$ but not to $N_{1,2,3\cdots m+1}$, this would be impossible
because the latter form a cover of the former. In any case as in the derivation
of the Cantor set, we can put the above countable series of sub sets of sub covers in a
one to one correspondence with suitable sub intervals of a real interval
$(a,b)$.\\
{\large \bf{Case I}}\\
If $N_{1,2,3\cdots m} \to$ an element of the set $N$ as $m \to \infty$, that
is if the set is closed, we would be establishing a one
to one relationship with points on the interval $(a,b)$ and hence could use the
metric of this latter interval, as seen earlier.\\
{\large \bf{Case II}}\\
It is interesting to consider the case where in the above iterative countable
process, the limit does not tend to an element of the set $N$, that is set $N$
is not closed and has what we may call singular points. We could still truncate
the process at $N_{1,2,3\cdots m}$ for some $m > L$ arbitrary and establish a
one to one relationship between such truncated subsets and arbitrarily small
intervals in $a,b$. We could still speak of a metric or distance between two
such arbiitrarily small intervals.\\
This case is of interest because of recent
work which describes elementary particles as, what may be called Quantum
Mechanical Kerr-Newman Black Holes or vortices, where we have a length of the order of the
Compton wavelength as seen in the previous section, within which spacetime as
we know it breaksdown. Such cut offs as seen lead to a non commutative geometry
(\ref{e14}) and
what may be called fuzzy spaces\cite{r28,r29,r4}.(We note that the centre of
the vortex is a singular point). In any case, the number of particles in the universe is of
the order $10^{80}$, which approxiimates infinity from a physicist's point of view.\\
Interestingly, we usually consider two types of infinite sets - those with cardinal number $n$
corresponding to countable infinities, and those with cardinal number $c$
corresponding to a continuum, there being nothing inbetween. This is the well
known but unproven Continuum Hypotheses.\\
What we have shown with the above process is that it is possible to concieve an intermediate possibility
with a cardinal number $n^p, p > 1$.\\
We also note again the similarity with El Naschie's transfinite Cantor sets.
In the above considerations three
properties are important: the set must be closed i.e. it must contain all its limit
points, perfect i.e. in addition each of its points must be a limit point
and disconnected i.e. it contains no nonnull open intervals. Only
the first was invoked in Case I.
\section{The Generalized Uncertainity Principle}
In theories of Quantum Gravity and also String Theory we encounter what may be
called the Generalized Uncertainity Principle
\begin{equation}
\Delta x \geq \frac{\hbar}{\Delta p} + \alpha \cdot \frac{\Delta p}{\hbar}\label{e15}
\end{equation}
This is symptomatic of the non zero spacetime extension, and indeed also follows
from (\ref{e14}). This could be construed to imply a correction to the velocity
of light as has been noted in the literature. It could also be taken to be a
correction to the Einstein mass-energy formula $\sim 0(l^2)$ (Cf.ref.\cite{r9}
for details).\\
A more complete picture emerges from the following simple model of a one dimensional
lattice, the points being spaced a length $l$ apart. In this case the energy,
as is known, can be shown to be given by
\begin{equation}
E \sim mc^2 cos (\alpha l)\label{e16}
\end{equation}
where $\alpha$ is proportional to the wave number.\\
A comparison with results following from (\ref{e14}) or (\ref{e15}) shows
that the latter are truncated versions of (\ref{e16}), truncated $\sim 0(l^2)$.

\end{document}